\documentclass[seceq]{ptptex}

\usepackage{graphicx,color} 
\usepackage{amssymb} 
\usepackage{latexsym}

\title{
A Scalable Low-Cost-UAV Traffic Network (uNet) %
}

\author{Santosh \textsc{Devasia}\footnote{Professor, Mechanical Engineering Dept., Box 352600,  AIAA Senior Member.}
Alexander \textsc{Lee}\footnote{Volunteer Student, Mechanical Engineering Dept., Box 352600.}}

\inst{
U. of Washington, Seattle, WA 98195-2600, USA
}

\abst{
This article proposes a new Unmanned Aerial Vehicle (UAV) operation paradigm to enable a large number of relatively low-cost UAVs to 
fly beyond-line-of-sight 
without costly sensing and communication systems or substantial human intervention in individual UAV control. 
Under 
current free-flight-like paradigm, wherein a UAV can travel along any route as long as it avoids restricted airspace and altitudes.
However, this requires expensive on-board sensing and communication as well as substantial human effort in order
to ensure avoidance of obstacles and collisions.
The increased cost serves as an impediment to the emergence and development of broader UAV applications.
The main contribution of this work is to propose the use of pre-established route network for UAV traffic management, 
which allows: (i)~pre-mapping of obstacles along 
the route network to reduce the onboard sensing requirements and the associated costs for avoiding such obstacles; 
and (ii)~use of well-developed routing algorithms   to select UAV schedules that avoid conflicts.
Available GPS-based navigation  can be used to fly the UAV along the selected route and time schedule 
with relatively low added cost, which therefore, reduces  
the barrier to entry into new UAV-applications market.   
Finally, this article proposes a new decoupling scheme  for conflict-free transitions between edges of the route network 
at each node of the route network 
to reduce potential conflicts between UAVs and ensuing delays. 
A  simulation example is used to illustrate the proposed uNet approach.
}

\newtheorem{rem}{\textbf{Remark}}
\newtheorem{assump}{\textbf{Assumption}}

\begin{document}

\maketitle

\newpage

\section{Introduction}

This article proposes a new Unmanned Aerial Vehicle (UAV)  operation paradigm to enable a large number of relatively low-cost UAVs to 
fly beyond-line-of-sight,  
without costly sensing and communication systems or substantial human intervention in the individual UAV control. 
A broadly accessible UAV traffic network (uNet) for low-cost UAVs can reduce the barrier to entry into the UAV market and 
enable new commercial enterprises and applications. For example, just as mobile phones revolutionized communication in developing countries by circumventing costly landlines, minimalist UAVs could make product delivery a reality in geographic areas where transportation infrastructure is difficult and prohibitively expensive to build.
The uNet itself can be an agglomeration of new franchise operators, 
along with internet-based application (APP) providers who enable users to automatically negotiate between different operators, and get a UAV to its destination. 
Additionally, existing commercial delivery companies can use the uNet to transport products and smaller UAV brokerage firms can use it to transport 
UAVs between locations.


%

\vspace{0.1in}
Current approaches to UAV traffic management are based on the free-flight paradigm, where a UAV can travel along any route as long as it can avoid restricted airspace, and satisfy constraints such as altitude limits. Such freedom to fly anywhere, nevertheless, creates challenges when trying to avoid obstacles and collisions with potentially other UAVs.
 For example, without a-priori detailed information of the entire airspace, the free-flight approach leads to lack of knowledge about the  immediate surroundings of the UAV, which then requires sensors and/or humans to detect obstacles. 
The ultimate consequence of this is an increased cost of UAV operation, scaling upwards in complexity as more and more UAVs are introduced. 
Conflict resolution protocols in the free-flight system would need to continually evolve as density increases, 
and sensors and UAV-to-UAV communications would need to become progressively more sophisticated, e.g., 
for potential collaborative conflict resolution, as investigated previously for manned aircraft~\cite{Frazzoli01,Pallottino_02}. 
An approach that can avoid the increased complexity and cost of UAV-to-UAV communication 
is to develop pre-specified conflict avoidance rules for UAVs that are similar to visual flight rules (VFR)  for manned aircraft. 
However, it can be challenging to develop  such VFR rules  that are provably safe  with a large number of UAVs with multiple independent UAV operators. 
Another approach is to use human-guided conflict resolution  similar to  current air traffic management (ATM) for commercial manned flights over controlled airspace~\cite{Schmidt76,Janic97,erzberger95,Kenny_Thesis_2013}. 
The human effort required for such conflict resolution can be lessened by using emerging concepts such as 
sense and avoid~\cite{Mitre_UAV_collision_avoidance}. Nevertheless, the human workload for conflict detection and avoidance tends to increase with number of aircraft, which limits the ability 
to scale such human-centered methods for a large number of UAVs.

\vspace{0.1in}
The new idea proposed here is to place the UAVs along an established route network, similar to automated guided vehicles on a factory floor~\cite{Ling_02,Smolic_2010}, 
 or the jet routes followed by commercial 
 aircraft in controlled airspace~\cite{devasia_99atm}. 
 These routes can be dense and time-varying to optimize for and accommodate conditions such as wind speed, precipitation, 
 and other potential local variables. Routes could be ad-hoc networks set up to meet application requirements, e.g., 
for agricultural applications or coordinated disaster relief. 
 Even with a high density of UAVs in flight, such route networks can provide sufficient flexibility while also addressing privacy concerns. 
A UAV using this route network could get to a typical home  by 
flying over conventional roads,  without flying  over private property or other restricted areas.

\vspace{0.1in}
Compared to the free-flight-style of UAV traffic management, the proposed approach over 
established route networks (such as roadways) offers two major advantages. First, obstacles along the routes 
can be mapped a-priori and updated as needed, reducing onboard sensing requirements. 
Note that  detailed mapping of obstacles is only needed along the routes in this approach and 
not the entire airspace as in the free-flight approach.
Waypoints along these established routes can be used to fly the UAV along a three dimensional trajectory, e.g.,~\cite{Besada_Portas_2010,Yang2015atm}. 
Moreover, the initial arrival into and the final departure from the uNet could be managed by humans, or using GPS-based landing schemes, e.g.,~\cite{ChoUAVLanding_07,Carnes_Thesis_2014}.
Then, for UAV traffic management, the amount of sensing and data required on each UAV reduces to GPS-enabled navigation from a given waypoint to the next specified waypoint. 
Since this does not require active imaging and sensing of the surroundings, such an approach could alleviate some of the privacy concerns associated with UAVs with substantial sensing capabilities. 
 
\vspace{0.1in}
A second advantage is that conflict resolution over a route network can be transformed into a resource allocation problem, i.e., two UAVs cannot occupy the same section of a route at the same time. 
Additionally, transitions between different sections of the network can be designed to be conflict-free for reducing potential delays, e.g.,~\cite{devasia_11atm}.
Then, conflict-free route selection can  be solved using well-developed existing resource allocation procedures for multi-agent systems, e.g.,~\cite{ter_Mors,Bertsimas_flow_capacity_00,Ling_02,Waslander_06,Mukherjee_flow_capacity_09,Smolic_2010,Allignol_12}.  For example, scheduling 
can be done  using  context-aware route planning (CARP), where a new agent (i.e., the UAV) selects a route from a route network (e.g., selects a shortest path-length route~\cite{Yen_Kshortest_71}) and schedules the arrival time  such that the new agent  is conflict-free with respect to previous 
agents that have been scheduled already~\cite{ter_Mors}. 
This a-priori de-conflicting using scheduling  reduces the amount of on-board sensing and communication needed on each UAV. 
Thereby, 
the proposed uNet enables access to low-cost UAV, and consequently can lead to broad usage of the uNet.

\vspace{0.1in}
Another  contribution of this work is to propose a new decoupling scheme  for conflict-free transitions between edges of the route network 
at each node of the route network 
to reduce potential conflicts between UAVs. 
At low UAV densities, establishing a waiting protocol at nodes of the route network could be one possible method of conflict avoidance. 
Unfortunately, this method incurs 
additional fuel costs to slow down from cruise and hover, and then accelerate again. 
If the conflicts at the nodes can be avoided by design, such delays can be reduced or even eliminated entirely. 
This is similar to the design of multi-level interchanges 
at freeway  junctions that allow multiple traffic streams to pass through without crossing each other as opposed to the use of traffic lights or stop signs on typical  surface streets. 
%

\vspace{0.1in}
In addition to offering solutions for prior waypoint mapping and conflict resolution, 
the uNet approach is also capable of   distributed development and implementation through 
sector-level uNets  (sNets) that each manage UAVs inside their local region of the airspace. 
With minimal effort from the user, who only needs to specify initial and final locations along with an estimated time of arrival (ETA), a service, possibly a commercial web-application (APP), can select a route that spans multiple sNets and choose a scheduled time of arrival (STA) into the uNet based on route-availability. The APP serves to negotiate between different sNets  and manage different service fees and availability along en-route sNets to propose a  scheduled time of arrival (STA) into the uNet.
If the UAV meets requirements such as fuel for the flight, GPS and communication needs between the
UAV and sNets, and human-guided (or automated) initial and final transitions into and out of the uNet, then the flight is accepted into
the first sNet.
Communication between the sNet and a UAV, e.g., about waypoint specifications and current location updates (using GPS on the UAV),  can be done using cellular data. 
This allows each sNet to manage and monitor UAV flights and conflict avoidance in its airspace
In case of emergencies which is another 
important issue in beyond line of sight operation, e.g.,~\cite{Stevenson2015}, 
the sNet can potentially redirect the  UAV by providing new waypoints that could be precomputed for every section of the route network. 
Moreover, each sNet will keep track of route occupancy (resource allocation) in its airspace and dynamically update the available route network if needed.

\vspace{0.1in}
The proposed distributed  approach allows public-private partnerships to manage different aspects of the uNet such as management and regulation of the sNets. 
A progressive rollout and expansion of the uNet infrastructure, one sNet piece at a time, can enable organic growth of the overall uNet. Commercial groups 
(such as local supermarkets or malls) can develop and manage the 
local sNets, and sNets can be either refined over time into smaller sectors with a finer mesh or integrated together to form larger sectors. 
Similar to the development of the  National Science Foundation (NSF) NSFNET,  which provided the backbone communication service for the Internet,  the public sector could help the uNet  effort by developing backbone-type services, e.g., over highways to connect local sNets.

\vspace{0.1in}
The article begins with a description of the proposed uNet structure in Section~\ref{sec_proposed_uNet_structure}. 
The resource-allocation approach to UAV scheduling
is illustrated with a  simulation example in Section~\ref{sec_example}, which is followed by conclusions.

\vspace{0.1in}
\section{Proposed uNet structure}
\label{sec_proposed_uNet_structure}

\subsection{UAV routing through sNets}
\begin{figure}[!ht]
\begin{center}
\includegraphics[width=.8\columnwidth]{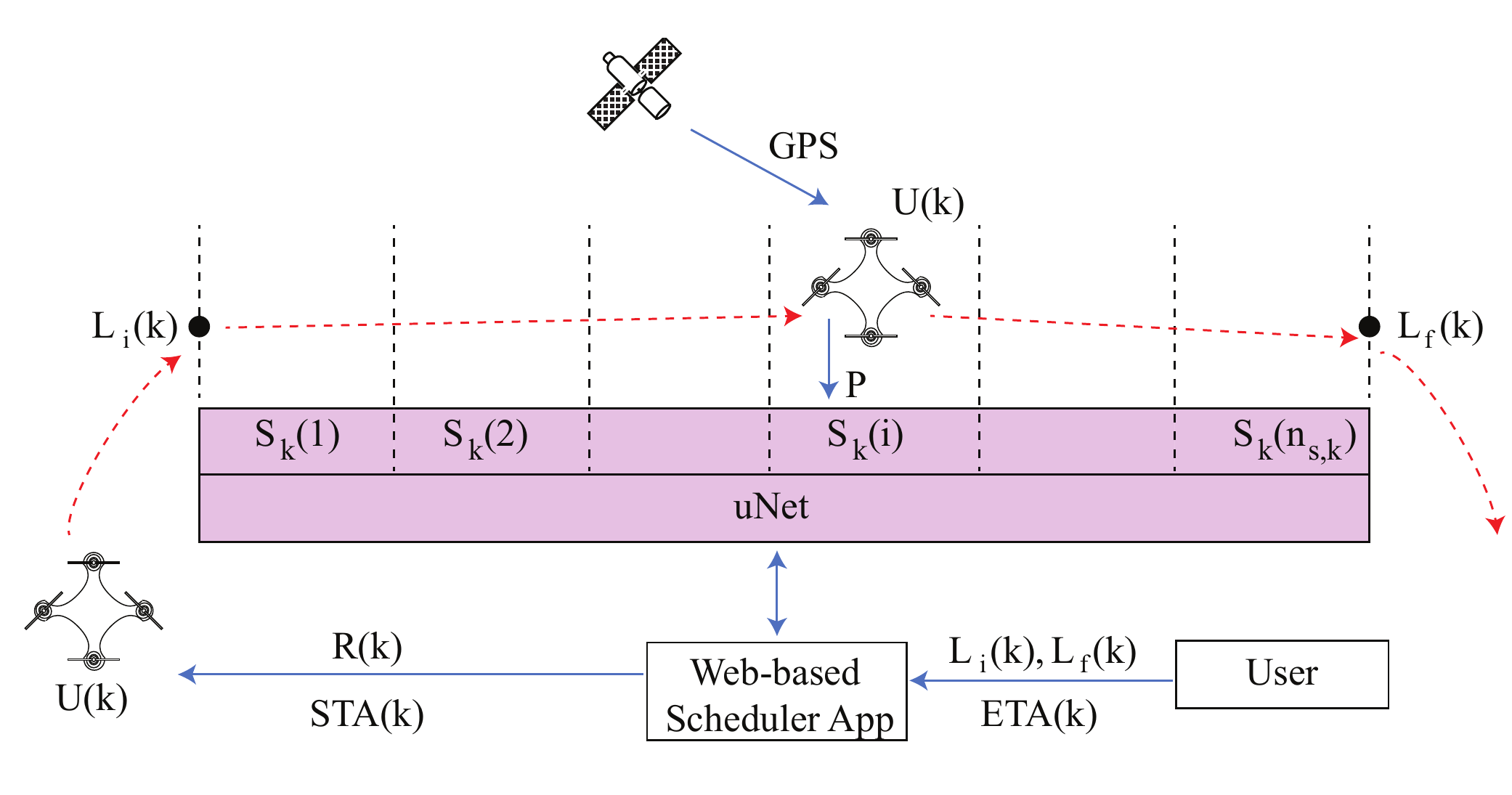}
\vspace{-0.1in}
\caption{Schematic routing of the $k^{th}$ UAV $U(k)$ through the uNet  from the initial location $L_i(k)$  in the initial sector-level uNet  $S_k(1)$  to  the final location  $L_f(k)$ in the 
final sector-level uNet  $S_k(n_{s,k})$. For a given expected time of arrival  ETA(k) into the uNet, the router negotiates  
between multiple sNets to determine the scheduled time of arrival  STA(k)  and a conflict-free flight route  $R(k)$ that spans multiple sNets  $\{S_k(i)\}_{i=1}^{n_{s,k}}$. 
The UAV  $U(k)$  receives GPS information and communicates its position  $P[U(k)](t)$ to the current sNet  $S_k(i)$ at time  $t$. 
}
\label{fig_schematic}
\end{center}
\end{figure}

The uNet consists of multiple sector-level uNets (sNets), each of which controls a specified region of the airspace. 
After entering a desired initial location $L_i$  in the initial sNet, a UAV $U$  can potentially transition through multiple 
such sNets before reaching the final location $L_f$  in the final sNet as illustrated in Fig.~\ref{fig_schematic}.

\suppressfloats

In particular, 
given the expected time of arrival  ETA(k) of the $k^{th}$ UAV  $U(k)$ into the initial location  $L_i(k)$ and the final location  $L_f(k)$, a router (potentially a web-based application) 
negotiates between multiple sNets  
to determine a scheduled time of arrival  STA(k)  and a flight route  $R(k)$ 
that is conflict-free with respect to all the prior UAVs  $[U(i)]_{i=1}^{k-1}$ in the uNet.

\vspace{0.1in}

\subsection{Definition: conflict-free}
In the following, two UAVs  $U(i)$ and $U(j)$  are considered to be in conflict at some time $t$ if they are inside the uNet region and their minimal separation is less than an acceptable value $d_{sep}$, i.e., 
\begin{equation}
\| L[U(i)](t) - L[U(j)](t) \|_2 < d_{sep} 
\end{equation}
where $\| \cdot \|_2$ is the standard Euclidean norm and $L [ U ( \cdot ) ] (t) $ is the position of the UAV  $U( \cdot ) $ at time  $t$. The minimal separation $d_{sep}$ is needed 
to ensure that there are no collisions with other UAVs,  even with potential errors in UAV positioning. Thus, the 
minimal separation $d_{sep}$ between UAVs,  needed to be 
be conflict-free, will depend on the precision of the GPS-based navigation system.

\vspace{0.1in}

The routes are selected by the local sNets to be obstacle free. 
Enroute sNets  $[S_k(i)]_{i=1}^{n_{s,k}}$ ensure that all routes including the selected  route $R(k)$ for the UAV $U(k)$ are clear of obstacles  
in each sNet  $S_k(i) \in {\cal{S}}$ that the $k^{th}$ UAV  $U(k)$ passes through where  ${\cal{S}}$ is the set of sNets and  $n_{s,k}$ is the number of sNets traversed as  in Fig.~\ref{fig_schematic}.
The flight route  $R(k)$ can be used to determine 
the fuel required for the flight.  If the UAV meets requirements such as fuel for the flight, GPS and communication needs between the UAV and sNets as well as human-guided (or automated) initial transition into and final transition from the uNet, then the flight is accepted by the first sNet  $S_k(1)$ on the selected flight route  $R(k)$. 

\vspace{0.1in}

\subsection{Specified route structure}
Under a specified route structure scenario, 
the potential  route  $R$ of a UAV  $U$ can be selected from edges of a  
directed graph  ${\cal{G}} = \left({\cal{N}}, {\cal{E}}\right)$ with nodes  
$ {\cal{N}}$ enumerated as  $\left[ N(i)  \right]_{i=1}^{n_{n}}$, $n_{n} >1$ 
and edges  $ {\cal{E}}  \subseteq {\cal{N}} \times {\cal{N}}$ enumerated as  
$\left[ E (j)\right]_{j=1}^{n_{e}}$, $n_{e}  \ge 1$. 
There is a path on the graph  from every desired initial location to every desired final location. 
It is assumed that each node in the uNet has a distinct spatial location, and the edges  connect distinct points in space. 
Therefore, the initial 
node  $N([j]_i)$ and the final node  $N([j]_f)$ of each edge  $E_{j}  = \{ N([j]_i),  N([j]_f) \}  \in {\cal{E}} $  are different, i.e., $ N([j]_i)\ne  N([j]_f) $.

\vspace{0.1in}

\begin{rem}[General 3-dimensional edges]
The edges  ${\cal{E}}$ between nodes in the graph  ${\cal{G}}$  could be general curved paths and can be three dimensional.
Alternatively, general curved paths could be approximated with straight-line segments and the start and end of these segments could be included in the set of graph nodes  $ {\cal{N}}$. 
\end{rem}

\vspace{0.1in}

Given an initial node  $L_i(k)$ and final node  $L_f(k)$ for the $k^{th}$ UAV  $U(k)$, 
a route  $R(k)$ 
is a path on the graph  ${\cal{G}} $, i.e.,   sequence of distinct edges, 
\begin{equation}
R_k= \left[ E_k(1), ~E_k(1), ~\ldots,  ~E_k(n_{r,k}) \right], \quad {\mbox{where}}~ E_k(j)  =   \{ N_k([j]_i),  N_k ([j]_f) \}   \in {\cal{E}},   \quad 
 \label{eq_route_defn}
\end{equation}
$n_{r,k}$ is the number of edges in route  $R_k$, which a sequence of distinct nodes 
\begin{equation}
N(R_k) =  \left[  L_i(k),  ~N_k ([1]_f), ~N_k ([2]_f),~ \ldots ,   ~N_k ([n_{r,k}]_f) \right].
 \label{eq_route_nodes_defn}
\end{equation}
Therefore,  the edges are connected, i.e., 
initial nodes  $N_k([\cdot]_i) \in {\cal{N}}$ and final nodes  $N_k([\cdot]_f) \in {\cal{N}}$ 
satisfy 
\begin{equation}
N_k([j]_f)= N_k([j+1]_i),  \quad \forall j = 1, 2,  \ldots, (n_{r,k}-1), 
 \label{eq_route_defn_edge_connectivity}
\end{equation}
without retracing an edge, i.e., 
\begin{equation}
N_k([j]_i) \ne  N_k([j+1]_f),  \quad \forall j = 1, 2,  \ldots, (n_{r,k}-1), 
 \label{eq_route_defn_edge_connectivity2}
\end{equation}
and the edges begin and end at the desired initial and final locations, i.e., 
\begin{equation}
N_k([1]_i) = L_i(k) ,  \quad N_k([n_{r,k}]_f) = L_f(k) 
 \label{eq_route_defn_edge_connectivity3}
\end{equation}

\vspace{0.1in}

\subsection{sNets}
\begin{figure}[!ht]
\begin{center}
\includegraphics[width=.45\columnwidth]{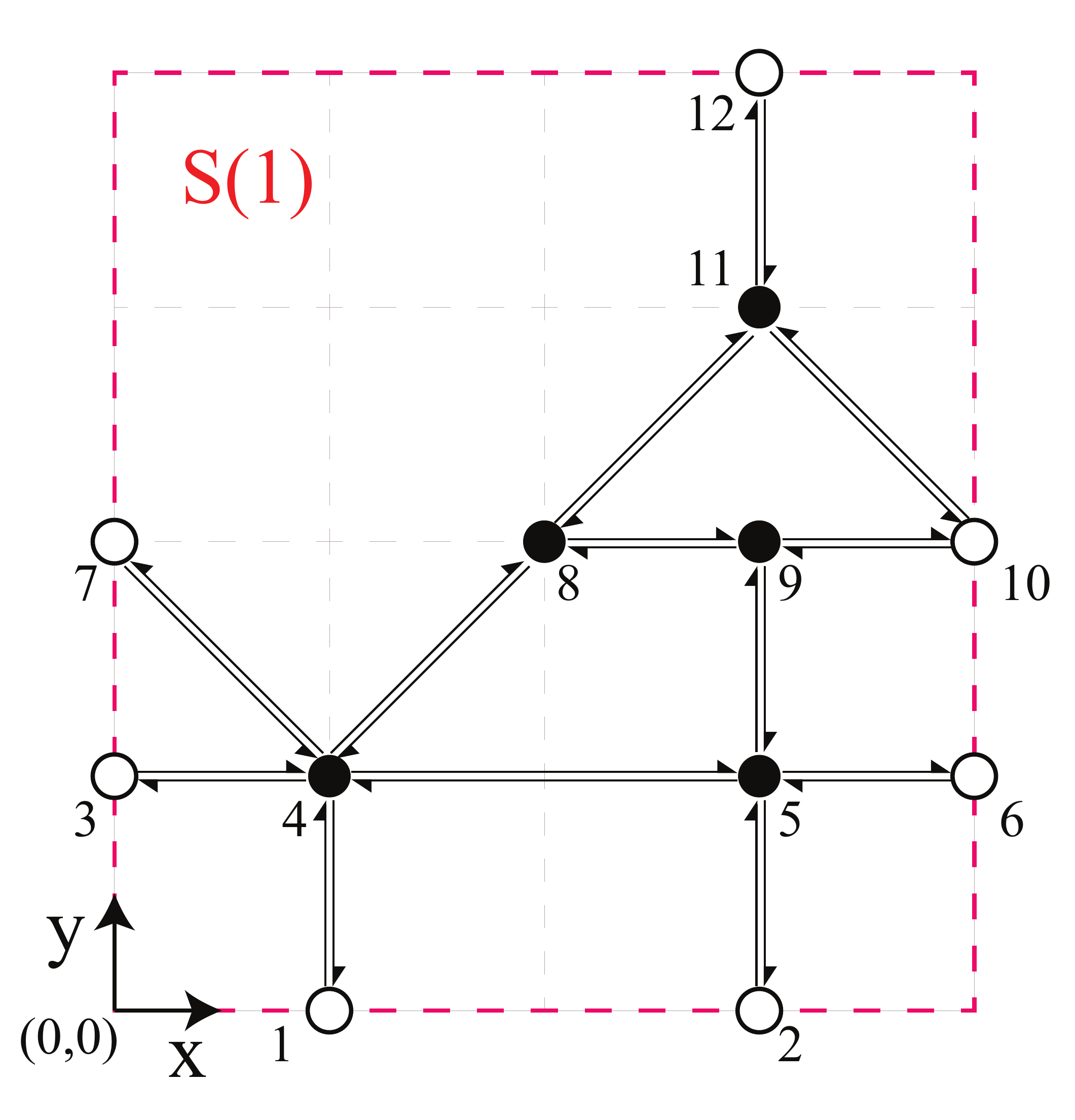}
\vspace{-0.1in}
\caption{Example sNet  $S(1)$ where nodes  ${\cal{N}}_1$ are represented by numbered dots on a square gird with the same vertical height. 
It includes  typical roadway 
intersection geometries such as $T$-intersection at node $9$, $Y$-intersection at node $8$, cross-intersection at node $5$, 
and a general non-circular intersection with more than four legs at node  $4$. 
}
\label{fig_sNet_1}
\end{center}
\end{figure}

The uNet 
is partitioned into $n_s$ number of sNets  $ {\cal{S}} = \left[S(j)\right]_{j=1}^{n_{s}} $ such that each edge inside the set of edges  ${\cal{E}}$ of the uNet graph  ${\cal{G}}$  belongs to a unique sNet. 
Each  sNet  $S(i)$ can be considered as a subgraph   
${\cal{G}}_i  = \left({\cal{N}}_i, {\cal{E}}_i\right) $  
of the overall  graph  ${\cal{G}} $ with $ {\cal{N}}_i \subseteq {\cal{N}}$  and $ {\cal{E}}_i \subseteq {\cal{E}}$. 
An example  
sNet  $S(1)$ is shown in Fig.~\ref{fig_sNet_1}, where the 
set of nodes  $ {\cal{N}}_1 \subseteq  {\cal{N}}$  associated with the sNet    
\begin{equation}
 {\cal{N}}_1 ~ =  \left[  N_1(1),  ~N_1(2),~ \ldots ,   ~N_1(12) \right] 
\end{equation}
is represented by numbered dots $\left[ i \right]_{i=1}^{12} $ on a square gird 
and the set of edges  
\begin{equation}
{\cal{E}}_1   ~~~   \subseteq \quad  \left ( {\cal{N}}_1 \times {\cal{N}}_1 \right)   \cap   {\cal{E}} 
\end{equation}
associated with the sNet is enumerated as $\left[E_1(i) \right]_{i=1}^{26}$ and given by 
\begin{equation}
 {\cal{E}}_1 = \left[ 
 \begin{array}{l}
 \{1,4\}, \{2,5\}, \{3,4\}, \{4,1\}, \{4,3\}, \{4,5\}, \{4,7\}, \{4,8\}, \{5,2\},  \{5,4\},
 \\
  \{5,6\}, \{5,9\},
  \{6,5\}, \{7,4\}, 
 \{8,4\}, \{8,9\}, 
 \{8,11\}, \{9,5\}, \{9,8\}, \{9,10\},
 \\
  \{10,9\}, \{10,11\}, 
 \{11,8\}, \{11,10\}, \{11,12\}, \{12,11\}
 \end{array}
   \right] .
\end{equation}

\suppressfloats

\subsection{De-conflicting UAV routes that do not share an edge}

Potential conflicts between two UAVs on two routes are classified into two scenarios: (i)~when the two routes do not share an edge; and 
(ii) when the two routes  share an edge. De-conflicting UAV routes that do not share an edge is discussed first. 
In particular,  it is 
assumed  in the following  that the uNet design is such that edges are sufficiently spaced from each other to avoid  
conflicts between two UAVs on two distinct routes that do not share any edges, 
especially far away from common nodes, as stated formally below. 

\vspace{0.1in}

\begin{assump}[Sufficiently-spaced edges]
\label{conflictfree_edge_assumption}
The overall uNet graph  ${\cal{G}}$ is constructed (with edges sufficiently spaced from each other) 
such that two UAVs one on each of two distinct edges, say $E(i)$ and $ E(j)$, 
cannot not have conflicts if the two edges do not have a  node  $N$ in common. 
\end{assump}

\vspace{0.1in}

\begin{assump}[Conflict-free away from nodes]
\label{conflictfree_far_from_Node}
If  two distinct edges  $E(i)$ and $ E(j)$  have a common node  $N$, then 
there is no conflict between two UAVs (one UAV on each of these two edges), 
when at least one of the UAVs is not inside a vertical conflict-free boundary cylinder  of radius $d^*[N]$ 
\begin{equation}
d^*[N] \ge d_{sep}
\label{Eq_boundary_sphere_distance}
\end{equation}
centered around the 
common node  $N$ and height $h^*[N]$.
\end{assump}

\vspace{0.1in}

\begin{assump}[Edges planar near nodes]
\label{assumption_intercepts_on_single_plane}
In the following, it is assumed that  for each node $N$ in the uNet graph  ${\cal{G}}$,  
each of the $n_N$ edges, say $E_{N}(i)$, in the set of edges  $ {\cal{E}}_N = \left[  {E}_{N}(j) \right]_{j=1}^{{n}_N}$ connected to the node $N$ 
intersects with the  conflict-free boundary cylinder    (in Assumption~\ref{conflictfree_far_from_Node}) at a single point $a_N(i)$. 
Moreover, the set of intersection points $ {\cal{A}}_N = \left[  {a}_{N}(i) \right]_{i=1}^{{n}_N}$ for each node $N$ lie on a single horizontal plane, on a conflict-free boundary circle 
$B_c[N] \subseteq  B_s[N]$.
\end{assump}

\vspace{0.1in}

\begin{figure}[!ht]
\begin{center}
\includegraphics[width=.75\columnwidth]{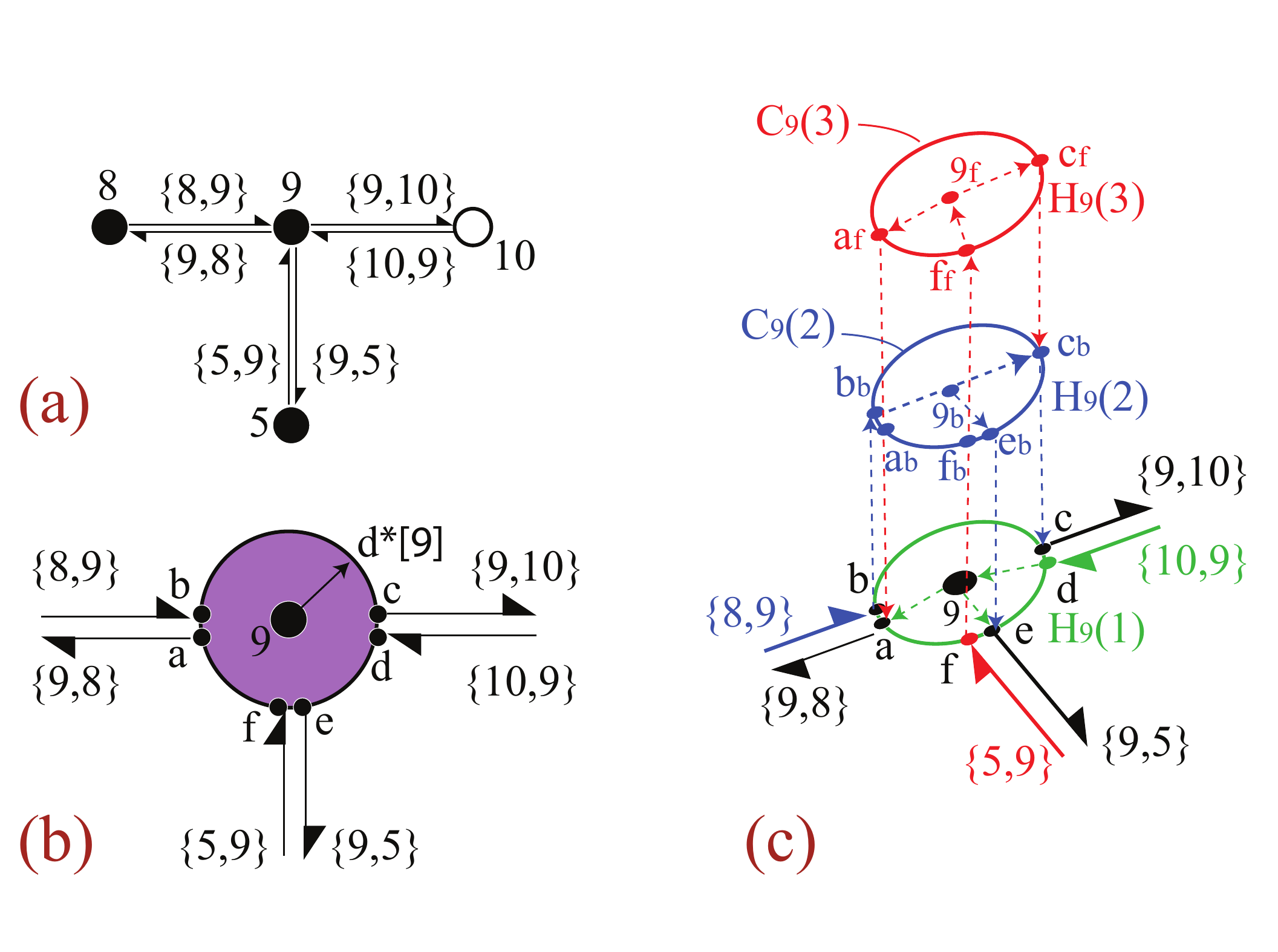}
\vspace{-0.1in}
\caption{Illustrative example for decoupling edge-to-edge transition at a node. 
(a) Example node $N=9$ of sNet $S(1)$ in Fig.~\ref{fig_sNet_1}. (b) Zoomed region around node $9$ illustrating 
conflict-free region, which is outside of the conflict-free boundary circle $B_c[9]$ of radius $d^*[9]$ centered around node $9$. (c)
Separating edge transitions into three levels  ${\cal{H}}_9 = \left[  {H}_{9}(j) \right]_{j=1}^{3}$ at different heights, one for each   edge into the  node $9$: 
green level ${H}_{9}(1)$  for edge $\{10,9\}$; blue level ${H}_{9}(2)$  for edge $\{8,9\}$; red level ${H}_{9}(3)$ for edge $\{5,9\}$.
}
\label{fig_Gen_node_decoupling}
\end{center}
\end{figure}

Assumptions~\ref{conflictfree_far_from_Node} and \ref{assumption_intercepts_on_single_plane} imply that all the edge-intersection points in the set $ {\cal{A}}_N$
on the circumference of the conflict-free  boundary circle $B_c[N]$ of a node $N$ are sufficiently separated, i.e., 
\begin{equation}
d({\cal{A}}_N)  ~~= \min_{i \ne j} \left\{ \,  d(N_i, N_j) \mid  N_i \in {\cal{A}}_N,   N_j \in  {\cal{A}}_N  \, \right\}   \quad \ge d_{sep}.
\label{Eq_intersection_node_spacing}
\end{equation}
Nevertheless, two UAVs $U(i)$ and $U(j)$ on routes  $R_i $ and $R_j$ respectively 
that share no edges could encounter conflict as they approach a shared  node  $N$,  i.e., 
$N \in  \left\{ P(R_i) \cap P(R_{j}) \right\} $ where the set of nodes of a route  $ P(\cdot)$ is defined in Eq.~\ref{eq_route_nodes_defn}. 
To illustrate, in the example sNet  $S(1)$ in 
Fig.~\ref{fig_sNet_1}, 
a UAV, say $U(1)$ on route $R(1)$  from node $8$ to node $10$ 
that includes a transition from edge $\{8,9\}$ to edge $\{9,10\}$  in Fig.~\ref{fig_Gen_node_decoupling}(a) could have a conflict 
near the common node $9$ with a UAV, 
say $U(2)$ on  route $R(2)$  from node $10$ to node $5$ that includes a transition from  edge $\{10,9\}$ to edge $\{9,5\}$.
When both UAVs are inside the space circumscribed by the boundary circle $B_c[9]$  shown in Fig.~\ref{fig_Gen_node_decoupling}(b) that is centered 
around node $9$, the potential for conflict needs to be anticipated and resolved.

\suppressfloats

\vspace{0.1in}

Potential conflicts,  between UAVs on routes that do not share an edge but share a common node as described above, can be avoided by 
sufficiently separating the  transitions between 
edges at all nodes. To achieve this separation,  the 
transitions from each edge in the set of edges  
$ \underline{\cal{E}}_N = \left[ \underline{E}_{N}(j) \right]_{j=1}^{\underline{n}_N}$
into a node  ${N}$ to an edge in the 
set of edges 
$ \overline{\cal{E}}_N = \left[ \overline{E}_{N}(j) \right]_{j=1}^{\overline{n}_N}$  out of the same node  ${N}$
are accomplished at  different-height levels 
$ {\cal{H}}_N = \left[ {H}_{N}(j) \right]_{j=1}^{\underline{n}_N}$.
Each level, e.g., ${H}_{N}(i)$ is a horizontal plane separated from all other levels by at least height $h_N$.
Let the UAV be transitioning from edge, say $\underline{E}_{N}(i)  \in \underline{\cal{E}}_N$, into 
node $N$ to edge, say  $ \overline{E}_{N}(j)  \in \overline{\cal{E}}_N$, out of node $N$. 
Moreover, let the intersections of these edges $\underline{E}_{N}(i)$ and $\overline{E}_{N}(j)$ with the 
the  conflict-free boundary circle  $B_c[N]$  (in Assumption~\ref{assumption_intercepts_on_single_plane}) be $\underline{a}_{N}(i)$ and 
$\overline{a}_{N}(j)$, respectively. Once the UAV on edge ${E}_{N}(i)$  reaches $\underline{a}_{N}(i)$ at the edge of the 
conflict-free boundary circle  $B_c[N]$,   
the UAV moves vertically to the $i^{th}$ level  ${H}_{N}(i)$, then horizontally on this level  ${L}_{N}(i)$ to be located directly above node $N$ 
before moving horizontally towards 
the location on level $i$ that is directly above $\overline{a}_{N}(j)$.  Then, the UAV descends vertically down from level  ${H}_{N}(i)$  to $\overline{a}_{N}(j)$ at the edge of the conflict-free boundary circle  $B_c[N]$  
and moves out of node $N$ along  the edge  $\overline{E}_{N}(j)$. 

\vspace{0.1in}

\begin{rem}[Transition levels above and below node]
The levels associated with the edge-transitions of a node $N$ need not be all above the node, they can be below or at the same height as the node $N$ provided 
such spaces are free of nodes, edges,  obstacles, or other constraints preventing use by the UAV. 
\end{rem}

\vspace{0.1in}

To illustrate the de-conflicted edge transitions, consider the node $N=9$ in the example sNet  $S(1)$, which has six edges, 
\begin{equation}
\begin{array}{rcl}
 {\cal{E}}_9 & = & \left[ E_9(1), ~ E_9(2), ~E_9(3), ~E_9(4), ~E_9(5), ~E_9(6) \right] 
 \\
 & = &  \left[ \{10,9\}, ~\{8,9\}, ~\{5,9\}, ~ \{9,10\}, ~ \{9,8\}, ~\{9,5\}  \right]
 \end{array}
   \label{All_Edges_node_9}
 \end{equation}
 connected to the node, 
where three of these edges, 
\begin{equation}
 \underline{\cal{E}}_9 = 
 \left[\underline{{E}}_9(1), ~\underline{{E}}_9(2),~ \underline{{E}}_9(3) \right]  = 
 \left[ \{10,9\}, ~\{8,9\}, ~ \{5,9\} \right]
   \label{Edges_into_node_9}
 \end{equation}
 are into   node $N=9$,  as shown in Fig.~\ref{fig_Gen_node_decoupling}(a). Therefore, 
three levels $ {\cal{H}}_9 = \left[  {H}_{9}(j) \right]_{j=1}^{3}$ (at different heights) are used for transitions from each of the edges in $\underline{\cal{E}}_9$ 
 into the  node $N=9$ 
 to the output edges  
 \begin{equation}
  \overline{\cal{E}}_9 =  
  \left[\overline{{E}}_9(1),~ \overline{{E}}_9(2),~\overline{{E}}_9(3) \right] = 
  \left[ \{9,10\}, ~ \{9,8\}, ~\{9,5\}  \right]
  \label{Edges_outof_node_9}
 \end{equation}
out of node $N=9$, as illustrated in  Fig.~\ref{fig_Gen_node_decoupling}(c).
Let a UAV aim to transition  from edge  $ \underline{E}_9(2) = \{8,9\} $ into node $N=9$ to edge $ \overline{E}_9(3) =  \{9,5\}$ out of node $N=9$.
Moreover, let the intersections of these edges $ \underline{E}_9(2)$ and  $ \overline{E}_9(2)$  with the 
the  conflict-free boundary circle $B_c[N]$  (in Assumption~\ref{assumption_intercepts_on_single_plane}) be $\underline{a}_{9}(2) = b$ and 
$\overline{a}_{9}(2) = e$, respectively,  as in 
Fig.~\ref{fig_Gen_node_decoupling}(c).   
Once the UAV on edge $\underline{E}_9(2) = \{8,9\}$  reaches $\underline{a}_{9}(2) = b$ on the circumference of the 
conflict-free boundary circle $B_c[9]$,  
the UAV moves vertically to the $2^{nd}$ level ${H}_{9}(2)$,  then horizontally on this level, 
${H}_{9}(2)$ to be located at $9_b$ directly above node $N=9$ 
before moving horizontally towards 
the location  $e_b$ on level ${H}_{9}(2)$ that is directly above the location $\overline{a}_{9}(2) = e$. 
Then, the UAV descends vertically down from level  ${H}_{9}(2)$  to location $\overline{a}_{9}(2) = e$ on the circumference of the conflict-free boundary circle  $B_c[9]$  
and moves out of node $N=9$ along  the edge  $\overline{E}_{N}(j) =  \{9,5\}$.

\vspace{0.1in}

\begin{rem}[Avoid back tracking]
Since the nodes in a route ($ P(\cdot)$ defined in Eq.~\ref{eq_route_nodes_defn}) are distinct, 
transitions to edges that return back to the start node of an edge into a node are not needed, e.g., 
transition from edge $\{8,9\} \in  \underline{\cal{E}}_9 $ to edge $\{9,8\} \in  \overline{\cal{E}}_9$ is not needed at node $N=9$, and is therefore, not shown in 
Fig.~\ref{fig_Gen_node_decoupling}(c).
\end{rem}

\vspace{0.1in}

Even with the multiple-level edge-to-edge transitions at a common node $N$, the minimal distance  between 
two UAVs on two different routes $R(i)$ and $R(j)$ without a common edge can be made larger than the acceptable value $d_{sep}$ to avoid conflicts. 
Note that, from Assumptions~\ref{conflictfree_edge_assumption}-\ref{assumption_intercepts_on_single_plane}, 
conflicts cannot occur on the edges if  both UAVs are outside the conflict-free boundary circle  $B_c[N]$  
of a common node, say $N$.  The UAVs have to arrive into the node $N$ through different edges since their 
associated routes do not share a common edge, and therefore,  are designed to achieve edge-transitions using different 
levels in the proposed scheme. Moreover, since the input and output edges from the node $N$  are different, the two UAVs also 
do not share the vertical ascending and descending paths. 

\vspace{0.1in}

\begin{figure}[!ht]
\begin{center}
\includegraphics[width=.4\columnwidth]{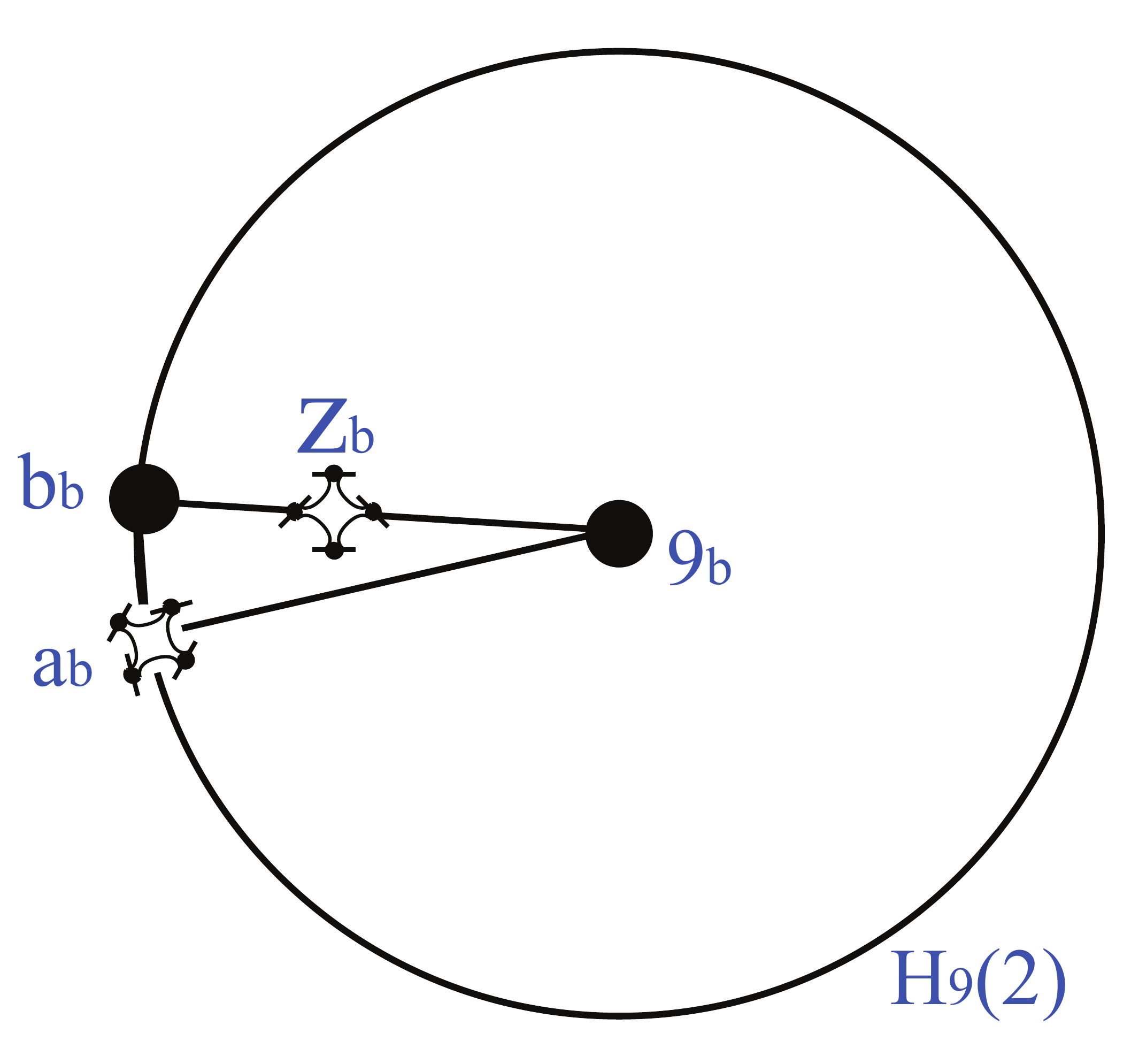}
\vspace{-0.1in}
\caption{Distance between two UAVs when both are on the same edge-to-edge transition level $H_9(2)$ from Fig.~\ref{fig_Gen_node_decoupling}c, e.g., when one   
UAV is on a vertical path transitioning to another level  $H_9(3)$ and is presently located at $a_b$ and the other is located at $z_b$ on a transition path from
location $b_b$ on level $H_9(2)$ to the location $9_b$ on the same level, above the node $9$. 
}
\label{fig_Min_Dis_node_9_decoupling}
\end{center}
\end{figure}

There are three possible scenarios under which a conflict can occur between two UAVs: (i)~both UAVs are on the ascending or descending paths to different edge-to-edge transition levels  ${H}_{N}(\cdot)$; 
(ii)~both UAVS are in  two edge-to-edge transition levels  ${H}_{N}(\cdot)$; and (iii)~one UAV is on one of the ascending or descending paths and the other is 
on an edge-to-edge transition level  ${H}_{N}(\cdot)$. 
The spacing between the vertical ascending 
and descending paths is at least as large as the minimal spacing  $d({\cal{A}}_N)$ between 
the intersection points $ {\cal{A}}_N$ of a node $N$, which in turn is larger than the acceptable conflict-free separation  $d_{sep}$ from Eq.~\ref{Eq_intersection_node_spacing}.  
Furthermore, there can be no conflicts under the second scenario when both UAVs are on different levels if the edge-to-edge transition levels are separated by at least  the acceptable conflict-free separation $d_{sep}$. 
Under the third scenario, the distance between two UAVs can become smaller than the minimal spacing  $d({\cal{A}}_N)$ between 
the intersection points  $ {\cal{A}}_N$ of a node  $N$. For example, consider the distance between two UAVs when both are on the same edge-to-edge transition 
level, e.g., $H_9(2)$ from Fig.~\ref{fig_Gen_node_decoupling}c,  when one   
UAV  $U(1)$  is is presently located at $a_b$ on a vertical path and is transitioning to another level, e.g., $H_9(3)$ while the other UAV  $U(2)$  is located 
at $z_b$ on a transition path from
location $b_b$ on level $H_9(2)$ to the location $9b$ on the same level, above the node $9$, as shown in Fig.~\ref{fig_Min_Dis_node_9_decoupling}.

\suppressfloats

Conflict can occur when UAV $U(1)$ is at location $a_b$, which is the closest that UAV $U(1)$ gets to  location $z_b$ of UAV $U(2)$, i.e., 
where its vertical ascent path intersects the edge-to-edge transition level $H_9(2)$.
The distance  $d(a_b,z_b)$  between the two UAVs  $U(1)$ and $U(2)$ can be found from triangle $\Delta(z_b9_ba_b) $ using the law of cosines as, 
\begin{equation}
d(a_b,z_b)^2  ~ = d(z_b,9_b)^2  ~+ d(a_b,9_b)^2 ~-2d(z_b,9_b)d(a_b,9_b)\cos( \angle z_b9_ba_b)
 \label{dist_U1U2_1}
 \end{equation}
 where the distance $d(a_b,9_b)$ is the radius $d^*[9]$ of the conflict-free boundary circle  $B_c[9]$, i.e., 
 \begin{equation}
 d(a_b,9_b) = d^*[9].
 \label{dist_U1U2_1_dBS}
 \end{equation}
 If the angle $\angle z_b9_ba_b$ is obtuse, then the UAVs cannot have a conflict since, from Eqs.~\ref{dist_U1U2_1}, and 
 \ref{dist_U1U2_1_dBS}, 
 \begin{equation}
\begin{array}{rcl}
d(a_b,z_b)^2  & = &  d(9_b,z_b)^2  ~+(d^*[9])^2~-2d(z_b,9_b)(d^*[9])\cos( \angle z_b9_ba_b) \\
& \ge & \left( d^*[9] \right)^2 \quad {\mbox{since}}~  cos( \angle z_b9_ba_b) \le 0  \\
& \ge & \left( d_{sep} \right)^2 \quad {\mbox{from Eq.~\ref{Eq_boundary_sphere_distance}}}.
 \end{array}
 \label{dist_U1U2_2}
 \end{equation} 
When the angle $\angle z_b9_ba_b$ is not obtuse, i.e., 
\begin{equation}
 \angle z_b9_ba_b  ~ =  \angle b_b9_ba_b   ~  \le \pi/2, 
 \label{dist_U1U2_2_angle}
\end{equation}
the distance  $d(a_b,z_b)$ between 
 the UAVs cannot be smaller than the perpendicular distance $d_{\perp}(a_b,\overline{b_b9_b})$ between location $a_b$ and line segment $\overline{b_b9_b}$, 
 i.e., 
  \begin{equation}
  \begin{array}{rcl}
d(a_b,z_b) ~\ge d_{\perp}(a_b,\overline{b_b9_b}) & = &  d(a_b,9_b) \sin{(\angle b_b9_ba_b)} \\
& = &   d^*[9]  \sqrt{  1 -\cos^2{(\angle b_b9_ba_b}) }.
 \label{dist_U1U2_2_min}
 \end{array}
 \end{equation}
Moreover, from the law of cosines for triangle $\Delta(b_b9_ba_b) $, 
  \begin{equation}
  \begin{array}{rcl}
 \cos{(\angle a_b9_bb_b}) & = &  \frac{d(b_b,9_b)^2  ~+ d(a_b,9_b)^2 - d(a_b,b_b)^2}{ d(b_b,9_b)d(a_b,9_b) } \\
 & = &  \frac{2 (d^*[9])^2  - d(a_b,b_b)^2}{ 2  (d^*[9])^2 }  ~ =  1 -  \frac{ d(a_b,b_b)^2}{ 2  (d^*[9])^2 } \\
 \end{array}
   \label{dist_U1U2_2_min2}
 \end{equation}
\noindent 
since distances $d(a_b,9_b)$ and $d(b_b,9_b)$ are equal to the radius $d^*[9]$ of the conflict-free boundary circle  $B_c[9]$, i.e., 
\begin{equation}
d(a_b,9_b) ~ = d(b_b,9_b)~ = d^*[9]
\label{eq_sides_boundary_sphere}
\end{equation}
and  the law of cosines for 
triangle $\Delta(b_b9_ba_b) $, Eq.~\ref{dist_U1U2_2_angle} and Eq.~\ref{eq_sides_boundary_sphere} yield, 
 \begin{equation}
\begin{array}{rcl}
d(a_b,b_b)^2  & = &  d(b_b,9_b)^2  ~+ d(a_b,9_b)^2 ~-d(b_b,9_b)d(a_b,9_b)\cos( \angle b_b9_ba_b) \\
& = &  2(d^*[9])^2  -(d^*[9])^2\cos( \angle z_b9_ba_b) \\ 
& \le & 2(d^*[9])^2.
 \end{array}
 \label{dist_U1U2_5}
 \end{equation} 
Then, from Eqs.~\ref{dist_U1U2_2_min},  \ref{dist_U1U2_2_min2} and \ref{dist_U1U2_5}, the distance between the UAVs $d(a_b,z_b)$ satisfies 
  \begin{equation}
  \begin{array}{rcl}
d(a_b,z_b) & ~\ge &   {d^*[9]}   \sqrt{  1 -\left[  1 -  \frac{ d(a_b,b_b)^2}{ 2  (d^*[9])^2 }  \right]^2 } 
~ =  d(a_b,b_b)    \sqrt{  \left[   1  -   \frac{  d(a_b,b_b)^2}{ 4 (d^*[9])^2 }  \right]^2 } \\
& \ge & d(a_b,b_b)    \sqrt{  \left[   1  -   \frac{  (\sqrt{2} d^*[9])^2}{ 4 (d^*[9])^2 }  \right]^2 } ~=  \frac{1}{\sqrt{2}}   d(a_b,b_b) ~\ge 
 \frac{1}{\sqrt{2}}    d({\cal{A}}_9)
 \label{dist_U1U2_2_min3}
 \end{array}
 \end{equation}
The  minimal distance $d({\cal{A}}_9) $  in Eq.~\ref{Eq_intersection_node_spacing} between 
intersection points ${\cal{A}}_9$ of node $9$ can be designed to be  large by choosing a sufficiently large 
radius $d^*[9]$ of the conflict-free boundary circle $B_c[9]$. A large enough radius  $d^*[9]$  enables  sufficient separation between the intersection points ${\cal{A}}_9$ 
that 
need to be distributed on the circumference of the conflict-free boundary circle  $B_c[9]$,  i.e., 
\begin{equation}
 d({\cal{A}}_N)  ~\ge \sqrt{2} \, d_{sep}
  \label{dist_U1U2_2_min4}
\end{equation} 
to ensure that the distance $d(a_b,z_b)$ between the UAVs satisfies the separation requirement, 
\begin{equation}
d(a_b,z_b)  ~\ge     \frac{1}{\sqrt{2}}    d({\cal{A}}_N)   ~\ge   d_{sep}.
\end{equation}

\vspace{0.1in}

\begin{assump}[Sufficiently-spaced intersection points]
\label{assumption_separation_of_intercepts}
Each node $N$ that has more than one  edge-to-edge transition, i.e., 
a potential for conflict during  edge-to-edge transition,  the 
minimal distance $d({\cal{A}}_N) $  in Eq.~\ref{Eq_intersection_node_spacing} between 
intersection points ${\cal{A}}_N$ of node $N$ is sufficiently large, i.e., satisfies Eq.~\ref{dist_U1U2_2_min4}.
Moreover, the height $h^*[N]$ of the conflict-free boundary cylinder in Assumption~\ref{conflictfree_far_from_Node} is sufficiently large 
so that the edge-to-edge transition levels ${H}_{N}(\cdot)$ 
are sufficiently separated  ($h_N \ge  d_{sep}$) to avoid conflicts between UAVs at  different levels, and all levels 
can be contained inside the conflict-free boundary cylinder.
\end{assump}

\vspace{0.1in}

Under Assumptions~\ref{conflictfree_edge_assumption}-\ref{assumption_separation_of_intercepts} there are no conflicts between UAVs in the uNet whose routes do not share an edge.

\begin{figure}[!ht]
\begin{center}
\includegraphics[width=.95\columnwidth]{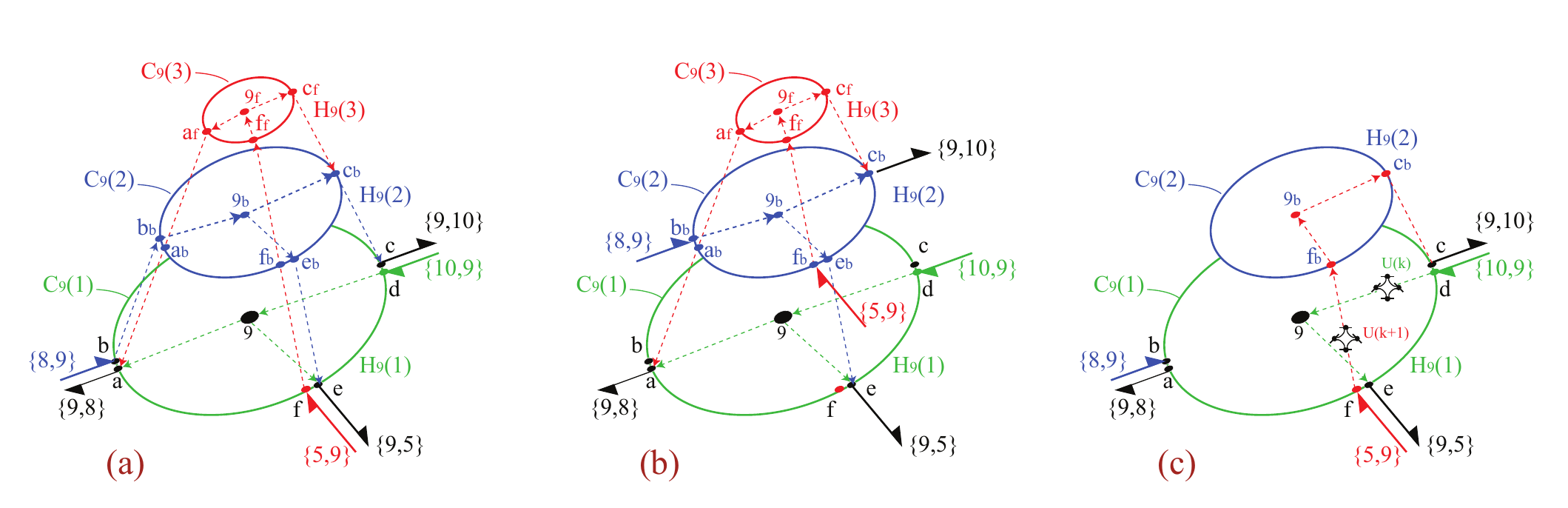}
\vspace{-0.1in}
\caption{Schematic of alternate approaches to  decoupling edge-to-edge transitions at node $9$. 
(a) Varying the radii of the conflict-free boundary circles (largest $c_9(1)$ to smallest $c_9(3)$) enables non-perpendicular changes 
to the different levels for edge-to-edge transitions as in Remark~\ref{rem_Non_vertical_ascent_descent}. 
(b) Accommodating  arrivals and departures at two different heights (levels) in narrow streets,  
as in Remark~\ref{rem_Narrow_streets}.
(c) On-demand conflict resolution by changing to a different level  when needed. 
UAV $U(k)$ does not change level but UAV $U(k+1)$, with potential conflict with UAV $U(k)$ changes to the closest available 
level $H_9(2)$  for edge to edge transitions, as in Remark~\ref{rem_On_demand_de_confliction}.
}
\label{fig_alternate_decoupling}
\end{center}
\end{figure}

\begin{rem}[Non-vertical ascent and descent]
\label{rem_Non_vertical_ascent_descent}
If vertical ascent and descent is not feasible, the edge-to-edge transition scheme can be modified with gradual ascent and descent. 
In particular, by making the size of the circles $C_N(\cdot)$ around which transition points into and out of each level are made  larger for levels that are closer to the node $N$. 
For example, the circle  $C_9(3)$ at level $H_9(3)$ for the example in Fig.~\ref{fig_Gen_node_decoupling}, 
around which transition points at locations $a_f, c_f$ and $ f_f$ are arranged, can be the smallest, followed by a larger level $H_9(2)$ and level $H_9(1)$ can be the largest, as illustrated in Fig.~\ref{fig_alternate_decoupling}(a).
\end{rem}

\vspace{0.1in}

\begin{rem}[Narrow streets]
\label{rem_Narrow_streets}
In narrow streets, for sufficient separation between edges, only one edge might be permitted, unless the UAVs fly above the buildings. 
Alternatively,  two edges could be stacked vertically one above the other to enable flight in two directions, 
In this case the arrivals can be separated into different levels, and the proposed approach can be used to ascend or descend to a different edge, as illustrated in Fig.~\ref{fig_alternate_decoupling}(b).
\end{rem}

\vspace{0.1in}

\begin{rem}[On-demand de-confliction]
\label{rem_On_demand_de_confliction}
The vertical ascent and descent adds time to the UAV flight. It is possible to only use the different levels when needed, e.g., as in on-demand de-conflicting schemes~\cite{Yoo14ondemand}. For example, if there are no conflicts at a node $N$  for the $k^{th}$ UAV $U(k)$ with previously scheduled UAVs, then the 
UAV $U(k)$ can directly proceed from the  intersection point  $\underline{a}_{N}(\cdot)$ into the node at  the circumference of 
the  conflict-free boundary circle  $B_c[N]$  to node $N$ followed by movement to the   intersection point  $\overline{a}_{N}(\cdot)$ out of the node $N$.  If there is potential for conflict, say for UAV $(U(k+1)$ with UAV $U(k)$ then the later UAV $U(k+1)$  selects the next-closest available level $H_N(\cdot)$ to  achieve the edge-to-edge transition, as illustrated in Fig.~\ref{fig_alternate_decoupling}(c).
\end{rem}

\suppressfloats

\vspace{0.1in}

\subsection{De-conflicted turns for UAV routes with shared edge}
\begin{figure}[!ht]
\begin{center}
\includegraphics[width=.4\columnwidth]{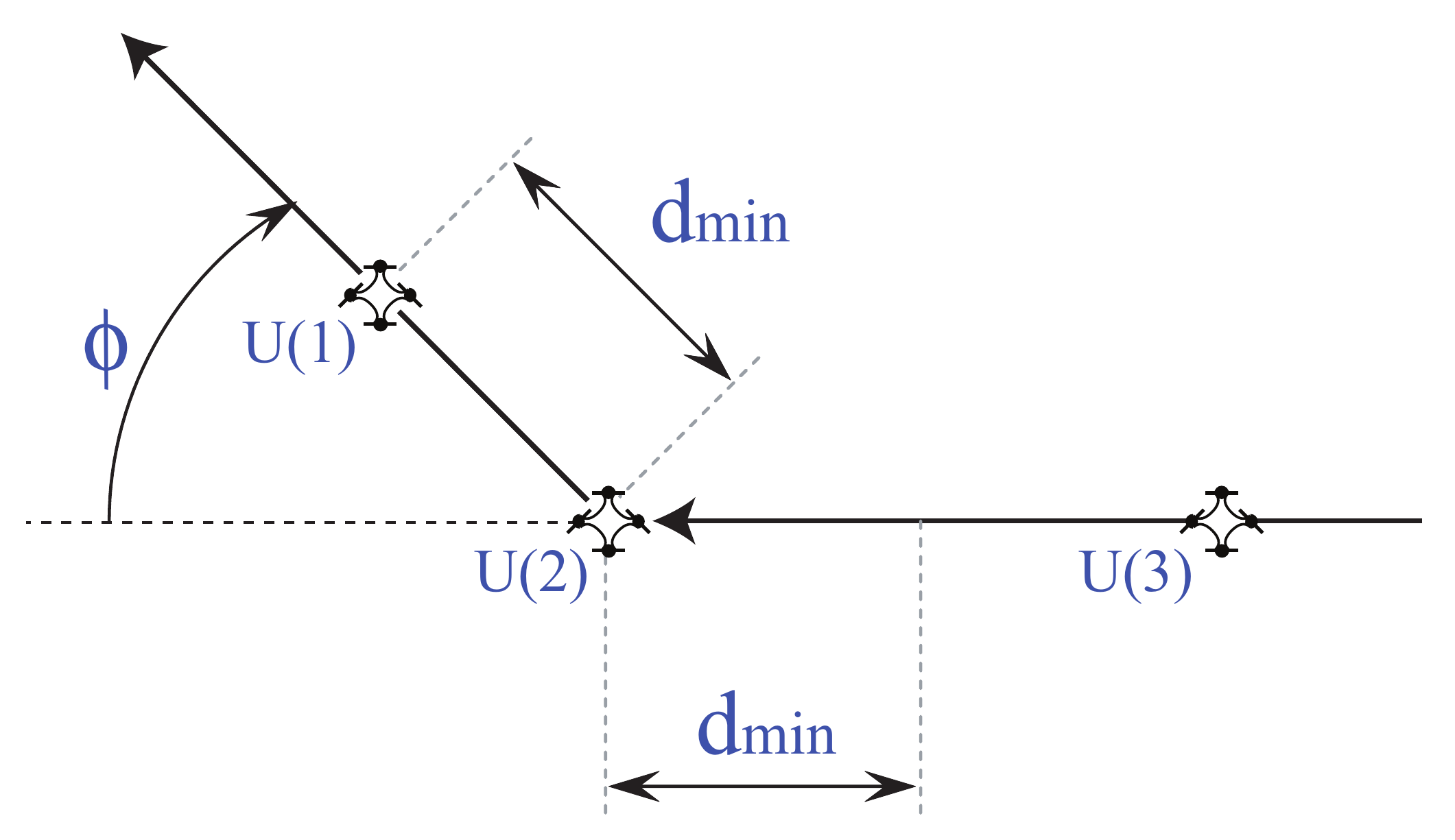}
\vspace{-0.1in}
\caption{UAVs, e.g., $U(1), U(2)$ and $U(3)$,  need to be spaced by at least the minimal distance $d_{min}$ in Eq.~\ref{eq_min_spacing} 
when following each other on a turn of angle $\phi \le \phi^*$ to avoid conflicts. 
}
\label{fig_Min_spacing_turns}
\end{center}
\end{figure}

The second scenario for potential conflict occurs between UAVs on routes that 
share an edge. There are two conflict cases under this scenario: (i)~during turns when UAVs are following each other; and (ii)~during merges into an edge. The first case of conflicts during turns is studied in this subsection. Note that even if two UAVs on a single edge are sufficiently separated initially, 
the spacing between them can decrease when making turns, e.g., 
for making edge-to-edge transitions at a node or on turns within a single edge. 
Therefore, the spacing between UAVs following each other on a turn edge needs to be sufficiently large to ensure conflict-free turns as shown in previous works for 
conflict resolution, e.g.,~\cite{devasia_11atm,Yoo14ondemand}. 
In particular, if two UAVs have the same speed  $V$ (m/s) on an edge, then two UAVs  arriving on the edge, one following the other, are conflict-free during turns on the edge if they are 
separated in arrival time at the start of the edge  by minimal time $T_{min}$ that satisfies, e.g., see Lemma~2 in~\cite{devasia_11atm},
\begin{equation}
T_{min} ~ = \frac{1} {V } \, d_{min} ~= \frac{1} {V}  \left[     \frac{d_{sep}} {  \cos(\phi^*/2)  }    \right] 
\label{eq_min_spacing}
\end{equation} 
where $\phi^*< \pi$ is the maximum turn angle between straight sections of the route and $d_{min}$ is the minimal separation between UAVs along straight sections of the route, 
as shown in Fig.~\ref{fig_Min_spacing_turns}.  
This is formally stated in the following assumption. 

\suppressfloats

\vspace{0.1in}

\begin{assump}[Constant-speed flight]
\label{assumption_constant_speed}
The edges and transitions between the edges 
are considered to be straight segments and turn between these segments of angle $\phi$ less that the maximum turn angle of $\phi^*$. 
The speed of  UAVs along all straight sections in the route network is a constant  $V$ m/s and scheduled time of arrivals (STAs) into each input location $L_i (\cdot)$ in the route network will be spaced at least by the minimal time $T_{min}$ for conflict-free turns in Eq.~\ref{eq_min_spacing}. 
Moreover, the height $h^*[N]$ of the conflict-free boundary cylinder in Assumption~\ref{conflictfree_far_from_Node} is sufficiently large 
so that the edge-to-edge transition levels ${H}_{N}(\cdot)$ 
are sufficiently separated  
\begin{equation}
h_N \ge  d_{min}
\end{equation}
to decouple potential conflicts due to multiple heading changes at  different levels~\cite{devasia_11atm}, 
and all levels 
can be contained inside the conflict-free boundary cylinder.
\end{assump}

\vspace{0.1in}

\begin{rem}[UAVs with different speeds]
The  spacing condition in Eq.~\ref{eq_min_spacing} for conflict-free turns can be generalized to include UAVs with different speeds and turn dynamics, e.g., see~ spacing conditions developed in~\cite{Yoo13cdc,Yoo13turn}. Other solutions, e.g., with variable speed UAVs, could include waiting for sufficient time  just before arriving at each node to avoid conflicts with UAVs passing through the node that were scheduled earlier. However, the resulting wait-related delay could be avoided by the proposed de-conflicting approach using out-of-plane edge-to-edge transitions. 
\end{rem}

\vspace{0.1in}

\subsection{De-conflicted scheduling for merging into a shared edge}
\begin{figure}[!ht]
\begin{center}
\includegraphics[width=.35\columnwidth]{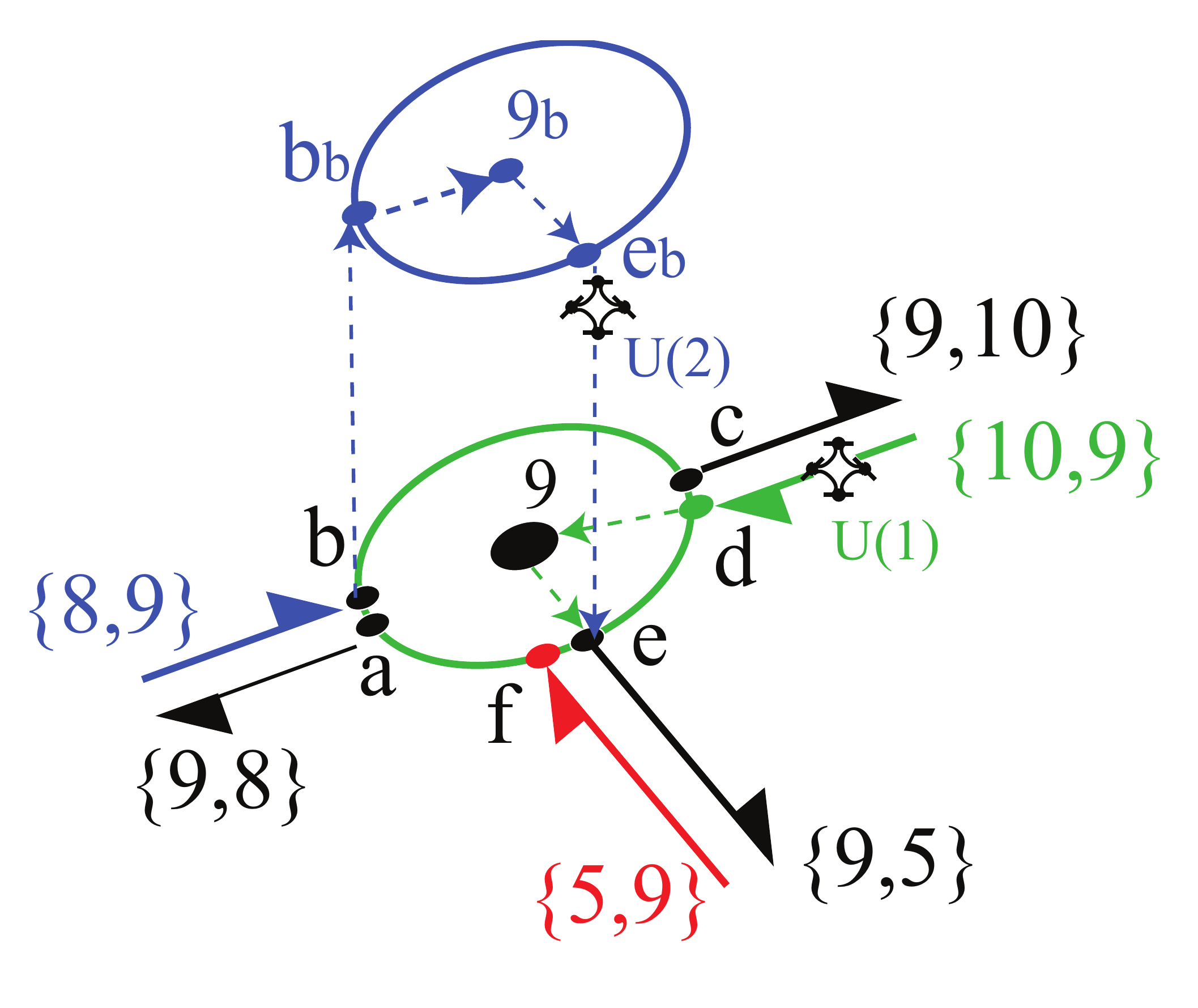}
\vspace{-0.1in}
\caption{Example merge into a shared edge, based on  Fig.~\ref{fig_Gen_node_decoupling}(c). 
Potential for conflict   when UAVs, say $U(1)$ and $U(2)$, transition  from two different edges  $\{10,9\} $ and  $\{8,9\} $ 
to the same edge $ \{9,5\}$ out of node $N=9$.
}
\label{fig_merge_conflict}
\end{center}
\end{figure}

A scheduling-based de-conflicting approach for the second case in the shared-edge scenario, i.e., for conflicts between UAVs  merging into a shared edge, is studied next.  
Consider two UAVs, say $U(1)$ and $U(2)$ transition from two different edges, say $\underline{E}_{N}(1)$ and  $\underline{E}_{N}(2)$, in the set of edges  $\underline{\cal{E}}_N$ into 
node $N$ to the same edge, say  $ \overline{E}_{N}(3)  \in \overline{\cal{E}}_N$, out of node $N$. Then,  
there is potential for conflict at the intersection point $\overline{a}_{N}(3)$ of the edge $\overline{E}_{N}(3)$ with the 
the  conflict-free boundary circle  $B_c[N]$.
For example, consider two UAVs $U(1)$ and $U(2)$ transitioning  from two different edges  $ \underline{E}_9(1) = \{10,9\} $ and  $ \underline{E}_9(2) = \{8,9\} $ 
into node $N=9$ to the same edge $ \overline{E}_9(3) =  \{9,5\}$ out of node $N=9$.
Then, there is potential for conflict at the intersection  $\overline{a}_{9}(3) = e$ of the output edge $ \overline{E}_9(3)$   with the 
the  conflict-free boundary circle  $B_c[9]$,  as in 
Fig.~\ref{fig_merge_conflict}. In particular, conflict  occurs if both UAVs $U(1)$ and $U(2)$ attempt to reach the location  $e$ at the same time.
Conflict can be avoided if the arrival times $t_e(1)$ and $t_e(2)$ for the two UAVs $U(1)$ and $U(2)$, respectively, at location $e$ of the edge $\{9,5\}$ out of the node $9$ 
are sufficiently separated, i.e., 
\begin{equation}
| t_e(1) - t_e(2)| ~~> ~d_{min}/V ~~= ~ T_{min}.
\end{equation}

\suppressfloats

\vspace{0.1in}

\begin{rem}[Sufficency of de-conflicting at arrival points]
Since the UAVs have the same travel speed $V$ from Assumption~\ref{assumption_constant_speed}, 
the  spacing of arrivals of UAVs  by the minimum time $T_{min}$  at any intersection point of an edge out of a node $N$ with its conflict-free boundary circle $B_c[N]$ 
ensures that the UAVs remain separated,  both before and after, as long as they continue to travel together along the same set of edges and nodes.
\end{rem}

\vspace{0.1in}

\begin{assump}[Isolation from UAVs outside uNet]
\label{assumption_conflict_free}
The uNet is sufficiently isolated and there are are no conflicts with UAVs in the uNet and other UAVs outside of the uNet, e.g., with UAVs 
prior to its arrival into the uNet at  an intersection point in the set $ {\cal{A}}_{L_i(k)} $ associated with  node $L_i(k) $
and  after its final departure from the UAV from an intersection point in the set $ {\cal{A}}_{L_f(k)}$ associated with node $L_f(k)$. 
\end{assump}

\vspace{0.1in}

At each intersection point ${a}_{N}(i)  \in {\cal{A}}_N $ of edges associated with node $N$ and its conflict-free boundary circle  $B_c[N]$, let the associated 
UAV arrival times be  $t_{{a}_{N}(i)}(k)$  such that $t_{{a}_{N}(i)}(k) < t_{{a}_{N}(i)}({k+1})$ for $ k < k+1$. 
Then, under Assumptions~\ref{conflictfree_edge_assumption}-\ref{assumption_conflict_free} and with the proposed de-conflicting scheme, 
there are no conflicts  provided the arrival times of UAVs $t_{{a}_{N}(i)}(k)$ (at each intersection point ${a}_{N}(i)$ of each node $N$)
are sufficiently separated, i.e., 
\begin{equation}
t_{{a}_{N}(i)}(k+1) - t_{{a}_{N}(i)}({k})   ~~ >  \, T_{min}.
\end{equation}

\vspace{0.1in}

\section{Example conflict-free UAV scheduling}
\label{sec_example}
The separation of arrival times at the intersection points $ {\cal{A}} =   \bigcup\limits_{N}  {\cal{A}}_N $ of edges and  conflict-free boundary circle  $B_c[N]$ of each node $N$ of the 
route network can be achieved using existing solutions to  resource allocation problems. 
Scheduling for the proposed uNet is illustrated with the 
context-aware route planning (CARP), where a new agent selects a route  and schedules the arrival time  
such that the new agent  is conflict-free with respect to previous agents~\cite{ter_Mors}.
Thus, CARP uses a  first come first served (FCFS) approach, which is not necessarily optimal over all agents, but is typically considered fair, e.g., in conventional Air Traffic Management~\cite{erzberger95}.

\vspace{0.1in}

\subsection{Example uNet}
\begin{figure}[!ht]
\begin{center}
\includegraphics[width=.75\columnwidth]{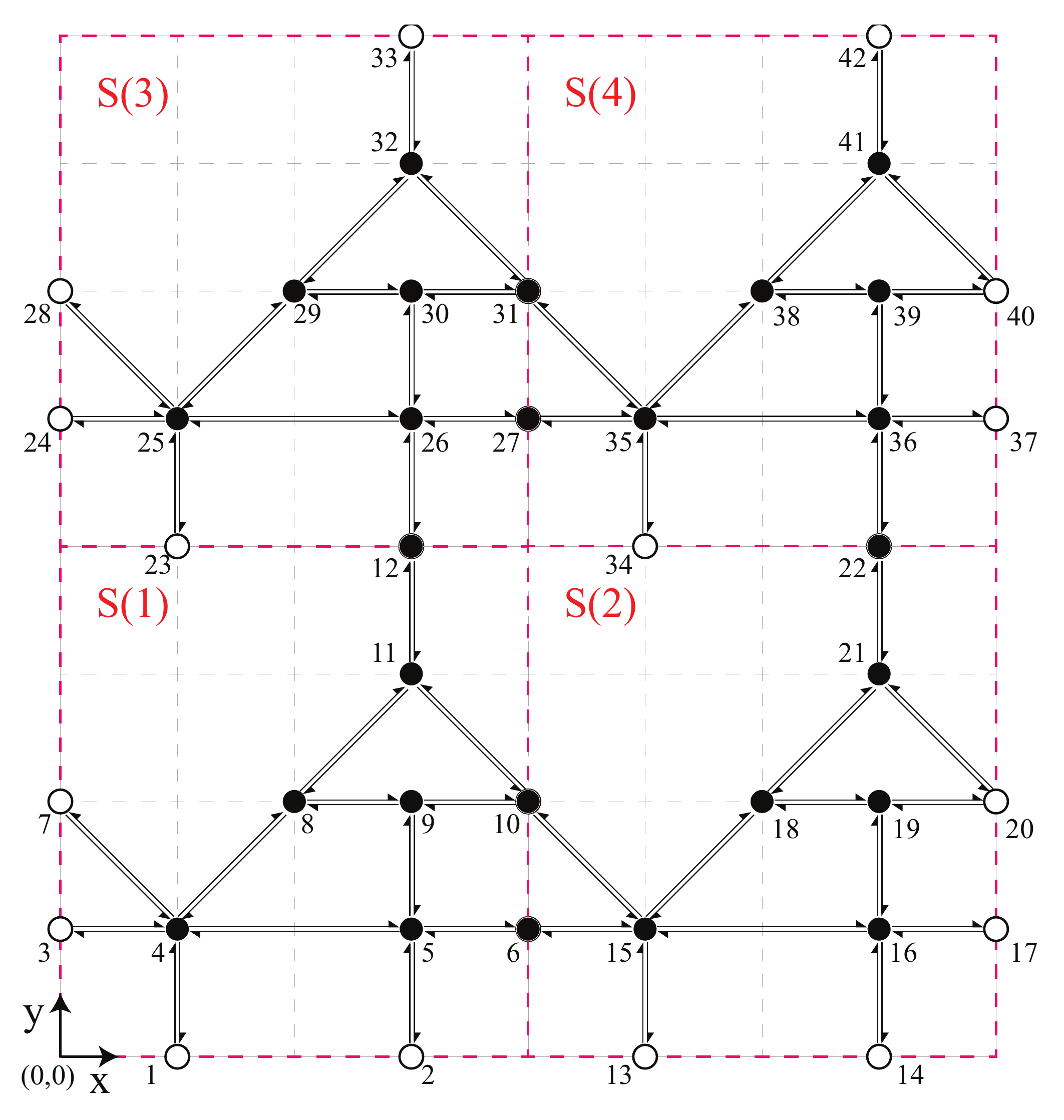}
\vspace{-0.1in}
\caption{Example uNet, with four sNets $S(1), S(2), S(3)$ and $S(4)$. Boundaries of the sNets are denoted by red dashed lines. 
Potential initial and final locations of UAVs on the uNet (e.g., node $1$) 
are depicted by white circles. 
}
\label{fig_uNet_example}
\end{center}
\end{figure}

Consider an example uNet, shown in Fig.~\ref{fig_uNet_example} composed of four repetitions of the sNet $S(1)$ in Fig.~\ref{fig_sNet_1}. 
Potential initial and final node  locations $L_i$ and $L_f$, respectively, are to be selected from the set ${\cal{L}}$ 
\begin{equation}
{\cal{L}} = \left\{ 1, 2, 3, 7, 13, 14, 17, 20, 23, 24, 28, 33, 34, 37, 40, 42 \right\} 
\end{equation}
depicted by white circles in Fig.~\ref{fig_uNet_example}. 
The grid spacing of the route network in Fig.~\ref{fig_uNet_example} is assumed to be $100$m, which is a medium-sized city block, e.g., in Seattle~\cite{Siksna_97}. 
Since GPS with Wide Area Augmentation System (WAAS) can have positioning error of  say $3-7$m, ideally the minimal separation $d_{sep}$ between UAVs should be $6-14$m. 
Given the five edge intersection at, say node $4$, 
de-conflicting the edge-to-edge transitions would require a separation of the UAV path into four paths at each level, e.g., as in Fig.~\ref{fig_Gen_node_decoupling}. Assuming that all 
the paths are uniformly distributed, i.e., at angle $2\pi / 5$ from each other, this requires a maximum heading change of 
$\phi^* = \pi - 2\pi / 5 = 3\pi / 5$. Then the minimal separation between the UAVs is  $d_{min} =  d_{sep}/ cos (\phi^*/2 ) = 10-23$m from Eq.~\ref{eq_min_spacing}. 
The spacing between UAVs on the same route is assumed to be $d_{min} = 20$m. 
With a speed of say $V= 4$m/s (about a tenth of the maximum expected UAV speed~\cite{FAA2015_UAV}), the time spacing $T_{min}$ 
between UAV arrivals then needs to be $T_{min} = d_{min}/V = 5$s.

\suppressfloats

\vspace{0.1in}

\subsection{Selection of UAV arrivals into uNet}
Simulations were performed to assess delays with the proposed FCFS scheduling approach. Note that the time needed to move across 
the eight grids of the example uNet in Fig.~\ref{fig_uNet_example}, with gridspacing of $100$m and speed $V=4$m/s  is $T_{8} = 25$s. 
Updates of the uNet, i.e., when new UAVs are aliowed to enter were performed at a time interval of $T_{\Delta} = 1$s  and 
time $t$ was discretized as 
\begin{equation}
t[n] =  n*T_{\Delta}.
\end{equation}
Let  a set of scheduled routes $R(i)$,  
$i = 1, 2, \cdots , k-1$ be given at  discretized time instant $t[n]$. 
Then a random number $r[n] \in [0, 1]$ was selected and a new UAV $U(k)$ was selected with an expected time of 
arrival $ETA(k) = t[n]$ if the random number was less than a probability of arrival $p_a$, i.e., 
\begin{equation}
r[n]  \le p_a.
\end{equation} 
The input  location  $L_i(k)$ and the final location $L_f(k)$ for the UAV $U(k)$ were chosen randomly from the set ${\cal{L}}$ 
Then,  the scheduled time of arrival  $STA(k)$ 
was found in a two step procedure: (i)~select a minimal distance route ${\cal{R}}(k)$ from the input location  $L_i(k)$ to the final location $L_f(k)$ through the network; 
and (ii) select  a conflict-free scheduled time of arrival $STA(k)$ 
\begin{equation} 
STA(k) \ge ETA(k), 
\end{equation} 
that is closest to the  expected time of arrival $ETA(k)$. 
The process of generating a random number at update time $t[n]$ was repeated, e.g., for UAV $U(k+1)$ until the random number was larger than the probability of a UAV arrival, i.e., 
$ r[n]  >  p_a$ resulting in no new UAVs. The update time was then incremented to $t[n+1]$. 
Note that more than one UAV can have an expected time of arrival ETA of $t[n]$. The process was stopped when the number of UAVs in the system reached $k=1000$.

\vspace{0.1in}

\subsection{Simulation results}
\begin{figure}[!ht]
\begin{center}
\includegraphics[width=.95\columnwidth]{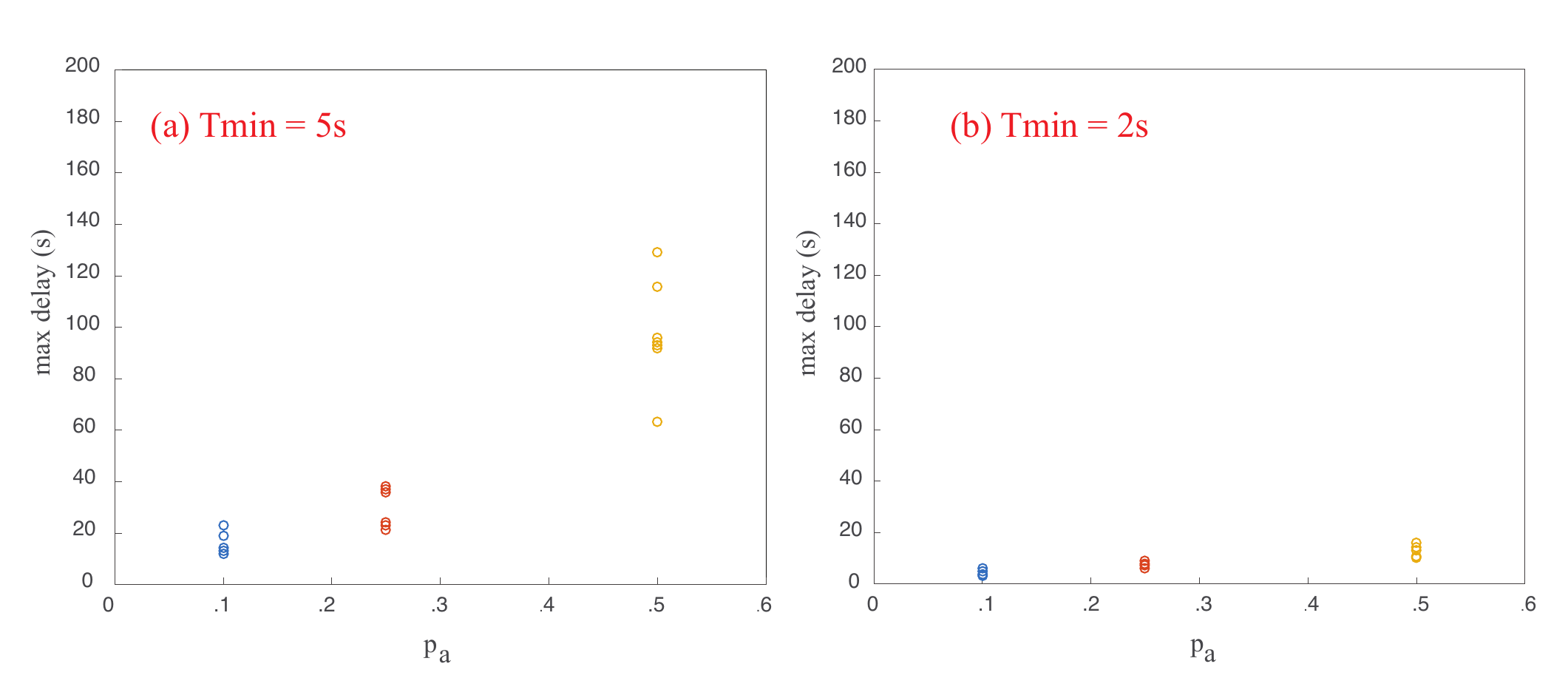}
\vspace{-0.1in}
\caption{
Maximum delays with seven simulation trials. Each trial simulation had 1000 UAVs and varying probability $p_a$. 
Two different arrival time separations  $T_{min}$ were used: left plot, $T_{min}=5$s; and right plot, $T_{min}=2$s; 
}
\label{fig_Max_Delay_7_Repetitions_1000_UAVs}
\end{center}
\end{figure}

The delay $D(k) = STA(k)  -ETA(k) $  for each UAV was found and is plotted as a function of the probability $p_a$ in Fig.~\ref{fig_Max_Delay_7_Repetitions_1000_UAVs} 
for a $1000$ UAVs. In these simulations, the time of flight from the  departure intersection point  in $ {\cal{A}}_N(i) $ of a node $N(i)$ to the arrival intersection point in 
$ {\cal{A}}_N(j) $  of another node $N(j)$ was approximated by 
the distance between the two nodes. 
As expected, the delays tend to increase with probability of arrival $p_a$. Delays can be reduced by using a smaller arrival time separation, e.g. for $T_{min}=2$s instead of 
for $T_{min}=5$s, as seen in Fig.~\ref{fig_Max_Delay_7_Repetitions_1000_UAVs}. 
The average maximum delay, e.g., for arrival probability $p_a = 0.5$ decreases by $87.3\%$, from $97.6$s to $12.4$s when the separation time is decreased from 
$T_{min}=5$s to $T_{min}=2$s.

\suppressfloats

\vspace{0.1in}

Substantial delays might not be acceptable to users. 
The simulation results indicate the effect of potential UAV conflict (e.g., the separation requirement) on delays. As the UAV density increases, or as the separation requirements
get large, there can be a substantial impact on the delays in the system. Clearly, one approach to reduce delays is to increase precision of the UAV navigation system, i.e., decrease the arrival time spacing $T_{min}$ between UAVs, however, this can increase costs. 
Alternate approaches are to choose a set of possible routes ${\cal{R}}(k)$ consisting of more than one viable route for each UAV. Then, an optimal route $R(k) \in {\cal{R}}(k)$ 
can be selected to optimize other criteria such as minimal delay between expected time of arrival at the destination ${ETA}_d(k)$ and scheduled time of arrival at the destination $STA_d(k)$. Note that a longer route (not energy optimal) might lead to a smaller delay at the destination. Other combinations of the energy cost (path length that can be approximated by ${STA}_d(k) - STA(k) $) and the delay at the destination ${STA}_d(k) - {ETA}_d(k)$ can be used to select a route $R(k)$ from the set of acceptable routes ${\cal{R}}(k)$.
Another approach to reduce delays is to modify the route network. If certain routes have high demand, additional routes could be added in parallel to these dense route or 
allow UAVs to bypass the congested areas. Similarly, direct higher-speed edges could be added between very high demand nodes to reduce the overall delay in the system.

\vspace{0.1in}

\subsection{Future work}
The implementation of the proposed uNet approach will require additional development and evaluation of protocols for emergencies. 
For example, If the UAV is not able to reach intermediate nodes in a specified time, e.g., due to uncertainties such as wind, 
routing is needed to land the UAV in a 
designated location, or the routing needs to be dynamically updated.  Such emergency re-routing could  be a pre-planned set of waypoints specific to each 
edge section in the route network.  Moreover, sensors on-board the UAV could also check for battery life  with planned stops for 
battery swaps for long-distance travel.  Moreover, to ensure that the UAVs can meet 
the uNet requirements, certification protocols need to be established and testing facilities are needed to prove UAV viability
before acceptance into the uNet.  Finally, the evaluation of the proposed uNet approach for human factors issues 
is needed to ensure that human supervisors can monitor and maintain overall situational awareness of overall system.

\vspace{0.1in}

\section{Conclusions}
This article proposed a new Unmanned Aerial Vehicle (UAV) operation paradigm to enable a large number of relatively low-cost UAVs to 
fly beyond-line-of-sight. The  use of an established route network for UAV traffic management was proposed to 
 reduce the onboard sensing requirements for avoiding such obstacles and enable the use of  of well-developed routing algorithms  
 to select UAV schedules that avoid conflicts.
 Another contribution of this work was to propose  a decoupling scheme  for conflict-free transitions between edges of the route network 
at each node to reduce potential conflicts between UAVs and ensuing delays. 
A simulation example and an example first-come-first-served scheduling scheme was used to illustrate the uNet approach.

\vspace{0.1in}

\section*{References}

\end{document}